# Physics of the Brain-Schizophrenia

*Probabilistic approach to consciousness and Hallucination*

OMID REZANIA [1,2]

[1] *York University, Department of Physics and Astronomy, Toronto, Canada*

[2] *Douglas Mental Health University Hospital, Aging Center, Montreal, Canada*

Correspondence: OMID REZANIA (omico80@yorku.ca)

Schizophrenic patients suffer from hallucination which its causality is not yet fully understood. This paper attempts to approach this mystery from the perspective of quantum mechanical theories. A novel approach has been adopted to demonstrate the hallucination as a time evolution of percepts basis states in the Hilbertian consciousness which are desynchronised from the time in real world. The method also extends his approach to predict mind and brain modulation through the correlation and coupling of consciousness and it reaches a clinically hypothetical outcome of inducing consciousness into a brain which is not conscious.

Schizophrenia is a complex neuropsychiatric disorder which is characterized by delusions, hallucinations, passivity phenomena, disordered thought process, disorganized behavior and progressive cognitive deficits. [1,5]

However, after many years of studying schizophrenia, the causality of this psychiatric disorder has not yet been fully understood.

There has been numerous publication to explain the causality of schizophrenia as the perturbation of consciousness from the quantum mechanical perspective, but no attempts have been made to address the causality of schizophrenia by exploiting the core nature of the consciousness, itself. [6.7].

The approach which has been proposed at this research has fundamentally addressed consciousness itself from a very novel perspective which not only explains the bizarre phenomena such as hallucinations but also provides a methodology which is capable to make predictions for future observations.

Consciousness regardless of its definition from any discipline can be represented as an infinite dimensional Hilbert space, where this novel approach to consciousness would play a significant role in explaining and predicting many yet unresolved mysteries in related scientific fields.

In quantum physics, Hilbert space is an abstract system where the rigorous mathematical formulation of dynamism of quantum systems is formulated. [8]

By this assumption, then we can liberally refer to consciousness as Hilbertian consciousness, where any percept can form a vector with unit length residing in that space, which would be referred to later as percepts basis vector inhabitant of the Hilbertian consciousness. These percepts now inhabitant of the Hilbertian consciousness following Dirac representation for quantum states: [8]

$$|percept>$$

However, every percept made by the individual at the same time forms and constructs a dual consciousness which is referred to as adjoint Hilbertian consciousness, represented similarly by the following notation:

$$<percept|$$

It can be well perceived that for any conscious or unconscious percept, both the percept basis state and its adjoint basis state are present simultaneously as a vector in the consciousness of individual, in other words any percepts represented as a basis vector in Hilbertian consciousness cannot exist individually without the coexistence of its adjoint in the adjoint Hilbertian consciousness.

In quantum mechanics, any quantum state represented as a vector in the Hilbert space can mathematically pair with an adjoint quantum state, not necessarily its own corresponding adjoint to extract a complex valued figure which has an important interpretation as a "probability amplitude". To extract that all important probability amplitude, basis vector states should mathematically satisfy orthogonality principle in the Hilbert space, which mathematically can be represented as $\delta - function$ which its value assumes either to be one or zero, and nothing else. [8]

However, when applied to quantum basis states, it essentially represents the functioning of an adjoint basis states on the quantum states to extract a probability amplitude (concept of probability amplitude is different from probability which would be hinted later). Quantum representation of orthogonality can be constructed in Dirac notation as [8]:

$$< n \mid m > = \delta(n, m)$$

$$where \ \ \delta(n, m) = 1 \ if \ n = m$$

$$\delta(n, m) = 0 \ \ if \ n \neq m$$

Above $m \ and \ n$ both represents any arbitrary quantum states which their inner product yields the probability amplitude of either one or zero, or simply states that the desired outcome would either happens 100% or it would never ever happen, and there is no other possibility in between.

In parallel, this approach assumes that when an individual consciously or unconsciously experiences a cognition, it stores that as percept which is represented as a vector basis state in his Hilbertian consciousness. Any percepts now an inhabitant of that complex space, couples with other percepts basis states but follows rigorously the orthogonality principle in analogous to Hilbert space inner products of two vectors.

As an example, the percept formed by smelling a flower and another percept similarly formed by seeing a picture of a cat, both form a basis vector and then they can couple by constructing an inner product in the Hilbertian consciousness of the individual who underwent those experiences. Later the mechanism of coupling of the percepts basis states between two different individuals are discussed and evaluated in detail.

According to the orthogonality principle outlined above these two percepts when couple and construct the inner product

$$< Smell \ of \ the \ flower | Picture \ of \ a \ cat > = 0$$

Produces the probability of zero.

The realisation of our outside universe is a matter of orthogonality of our percepts basis states already in our Hilbertian consciousness with the newly acquired percepts evolved in time.

This happens as an example in learning process. When an individual learn a French word such as *Bonne Journee* for the first time, he forms that as a percept basis state on his Hilbertian consciousness, which mathematically can be represented as $| \ Bonne \ Journee >$ alongside its simultaneous adjoint basis state formed in the adjoint Hilbertian state as $< Bonne \ Journee |$, then upon future hearing of that phrase, the only and only recognition happens when the newly heard of the phrase forming a new percept basis state on the Hilbertian consciousness, provided that the coupling or inner product of the trigger with the already formed adjoint basis state of the phrase be orthogonal as :

$$< Bonne \ Journee | Bonne \ Journee > = 1$$

It is obvious that the coupling of this basis state with a trigger which forms a newly vector upon hearing the same phrase in German yields zero due to the orthogonality of those basis vector in the Hilbertian consciousness

$$\langle\textit{Bonne Journe}|\textit{Guter Tag}\rangle = 0$$

Which zero probability amplitude of the above two basis states implies lack of recognition at the moment, but as German phrase already forms a permanent basis vector state on the individual`s Hilbertian consciousness and upon future hearing and due to the orthogonality, which yield one, the individual forms a recognition and memory shapes.

As it was implied then the formation of the memory can be interpreted as the complete sets of percepts formed as basis vector in the Hilbertian consciousness of the individual.

The picture which has been presented so far provides a static picture of the perception and cognition, which is defined only as coupling between two percepts formed as basis vector in the human Hilbertian consciousness and the element of the time which provides all crucial aspect of the temporality of the proposal has to be incorporated into the model.

The incorporation of time in this model sets the stage for an all-important boundary between hallucination and reality.

It is evident that outside universe is not static and it constantly evolves through the time, then our percepts basis states should also evolve in time to makes perception matches with outside events with respect to time.

But how temporal dimension of the Hilbertian consciousness can be incorporated with real experience events?

To visualise this scenario, a Clock can be imagined which is installed on your Hilbertian consciousness which measures the time for the evolution of the percepts basis states and a clock on the wall of your room.

For any percept formed due to an external event to be experienced in real-time, the two clocks, should be synchronised. This synchronisation manifest itself to experience real life events in real-time sequences.

Interestingly, if the time experience in the Hilbertian consciousness does not synchronise with the real-time events, then as a result the percepts basis vector states already inhabitant of Hilbertian consciousness evolve in time desynchronised with the real time. The result of these desynchronised time evolved percepts basis states in your Hilbertian consciousness is nothing but the hallucination which can be either experienced as dream or any symptoms of neurological or neuropsychiatric disorders such as schizophrenia or dementia with Lewy bodies.

However, this probabilistic approach to perception and cognition, can well be utilized from the quantum mechanical perspective to explain the memory and cognition impairment in other neurological disorders such as Alzheimer`s disease where the percepts basis vector states in the Hilbertian consciousness fails to correlate with the correct adjoint percept basis states in the Hilbertian consciousness of the patient , which in essence results in zero probability amplitude.

## Mind Modulation Through Consciousness Coupling

However, as any physics 'formulation should not only explain the physical phenomena, but also should be able to make some predictions about future applications. The model which has been presented here to explain the hallucination, a marked syndrome of schizophrenic patients can also be extended to predict the direct impact of Hilbertian consciousness of an individual on any other individual who they share some vector basis on their Hilbertian consciousness, which they correlate in the temporal dimension during which the basis vectors are constructed in their Hilbertian consciousness.

In this case, everyone if regarded in isolation in his Hilbertian consciousness has a complete superposition of all those cognitive experiences which they all exist with the same probability amplitude to occur at the same time before the triggering of an external occurrence which collapses the superposition only to one state which is your response.

Here, a completely new concept which is uniquely applicable to quantum mechanics is borrowed, probability amplitude $a$, which it relates to probability $P$ as $P = |a| * |a^*|$ which $a^*$ is the conjugate of $a$.[8]

In the simplest scenario, when an individual is exposed to some external stimuli, like smelling a flower or watching an accident or any other cognitive experience from external environment during a time interval which can span arbitrarily from 10 am to 11 am, we can consider the formation of a sub-Hilbertian consciousness where it can be constructed as a group formed by each cognitive experience which he has experienced limited to the time interval.

This sub-Hilbertian consciousness itself is an element of the Hilbertian consciousness of the individual.

For demonstration, the percept formed in the sub-Hilbertian consciousness would be represented as below, but due to limitation for graphical representation, I do represent each percept by a geometrical figure, in which case $|\blacksquare>$ may represent a percept formed by watching a picture of a cat now a basis vector in the sub-Hilbertian space, in its totality, then the complete sets of percept amplitude can form the following superposition in my consciousness:

$$|A> = \sum_1^n a1|\blacksquare> + a2|∴> + a3|\varDelta> + a4|\perp> + \cdots + an|\text{\textit{\#}}>$$

Where $a1, a2, a3, \ldots, an$ all refer to the probability amplitude of each percept basis states in the superposition constructed by them.

The individual superposition of percepts with equal probability amplitude formed during the time interval which is synchronised with other individual which is exposed to the same external stimuli during the same interval pairing with the other participant.

The percepts basis states can be completely different or similar but for the sake of reaching a sound conclusion we assume that some percepts are similar:

$$|B> = \sum_1^n b1|⚔> + b2|\blacksquare> + b3|⊾> + b4|\varDelta> + \cdots + bn|\text{\textit{\textflorin}}>$$

Which is similarly the complete superposition of percepts states in the sub-Hilbertian consciousness of the other individual.

The implication of the proposed model here comes with the idea of each participant`s consciousness which acts to collapse the complete superposition of percept basis states of the other participant in the pair, in other words each participant`s consciousness acts as if to measure the probability amplitude of the percepts of the other participant with respect to itself by forming the inner product as:

$$<B\,|\,A>$$

Which this product should conform to the orthogonality principle; [8]

$$<B\,|\,A> = (\sum_1^n b1^* <⚔| + b2^* <\blacksquare| + b3^* <⊾| + b4^* <\varDelta| + \cdots + bn^* <\text{\textit{\textflorin}}|)(\sum_1^n a1|\blacksquare> + a2|∴> + a3|\varDelta> + a4|\perp> + \cdots + an|\text{\textit{\#}}>)$$

Which can be summarized as:

$$b1^*a1 <⚔|\blacksquare> + b1^*a2 <⚔|∴> + b1^*a3 <⚔|\varDelta> + b1^*a4 <⚔|\perp> + \cdots + b1^*an <⚔|\text{\textit{\#}}> + \cdots + b4^*a1 <\varDelta|\blacksquare> + b4^*a2 <\varDelta|∴> + b4^*a3 <\varDelta|\varDelta> + b4^*a4 <\varDelta|\perp>$$

Due to the orthogonality principle then the consciousness to consciousness modality of a pair of individuals lead to the modulation of their mind with the following probability amplitude:

$$< B | A > = b2^* a1 + b4^* a3 + \mu$$

Where $\mu$ takes value from zero to any positive value depending on the similarity of mutual percepts of each individual.

Nonetheless, then the probability of finding the correlation and modulation of each brain to the brain of the other participants through the mutual consciousness correlation forming in the sub-Hilbertian space is:

$$P = (b2^* a1 + b4^* a3 + \mu)(b2^* a1 + b4^* a3 + \mu)^*$$
$$= (b2^* a1 + b4^* a3 + \mu)(b2 a1^* + b4 a3^* + \mu^*)$$

Which the probability with respect to the initial assumption with non-vanishing probability amplitudes for each percept basis states formed in the sub-Hilbertian consciousness of individual will never go to zero, which is an implication of the existence of the brain modulation through the correlation of the consciousness.

This provides a concrete analysis and a mathematical framework within the context of the quantum mechanics of the correlation of the consciousness between individuals who shares some percepts on their Hilbertian consciousness. This can be interpreted also as a brain modulation due to its consciousness correlation with others` due to the non-vanishing of the probability amplitudes of those percepts which formed a superposition state in Hilbertian consciousness.

But whether this method can be clinically adopted to induce consciousness to a patient who has lost consciousness leaves to be exploited further.

## COCLUSION

In this novel approach to consciousness, consciousness was regarded as an infinite dimension Hilbert space, referred to as Hilbertian consciousness where each conscious and unconscious experience and percept formed an inhabitant unit vector in that space. By carefully adhering to the orthogonality principle, cognition was interpreted as the time evolution of the percepts basis states triggered by the presence of the external stimuli. Hallucination as a characteristic symptom of schizophrenia then could be regraded as time evolution of the percept basis states which are not synchronised with real life temporal dimension, of all those conscious bases states which has been formed and constructed in the Hilbertian consciousness since the birth of the individual.

This approach has been adopted to make predictions which included the brain modulation through the correlation of the consciousness of the individuals who share some, although negligible percept basis vector states in their sub-Hilbertian consciousness. The approach was solely relied heavily on the principles of the probability amplitudes whose mod square produces the probability in its conventional scientific sense. It was shown that regardless of technical difficulties for realization of his approach, the brain modulation through correlation of the consciousness provides the probability which can never vanishes to zero.

## ACKNOWLEDMENTS

This research did not receive any specific grant from funding agencies in the public, commercial, or not-for-profit sectors.